\documentclass[preprint,superscriptaddress,prb]{revtex4}
\usepackage{graphicx}
\usepackage{amsmath}
\usepackage{units}
\usepackage[caption=false]{subfig}
\begin{document}
\title{Directed self-organization of graphene nanoribbons on SiC}

\author{M. Sprinkle}
\affiliation{Georgia Institute of Technology, Atlanta, Georgia
30332-0430, USA\\}
\author{M. Ruan}
\affiliation{Georgia Institute of Technology, Atlanta, Georgia
30332-0430, USA\\}
\author{X. Wu}
\affiliation{Georgia Institute of Technology, Atlanta, Georgia
30332-0430, USA\\}
\author{Y. Hu}
\affiliation{Georgia Institute of Technology, Atlanta, Georgia
30332-0430, USA\\}
\author{M. Rubio-Roy}
\affiliation{Georgia Institute of Technology, Atlanta, Georgia
30332-0430, USA\\}
\author{\\J. Hankinson}
\affiliation{Georgia Institute of Technology, Atlanta, Georgia
30332-0430, USA\\}
\author{N.K. Madiomanana}
\affiliation{Georgia Institute of Technology, Atlanta, Georgia
30332-0430, USA\\}
\author{C. Berger}
\affiliation{Georgia Institute of Technology, Atlanta, Georgia
30332-0430, USA\\}
\affiliation{CNRS/Institut N\'{e}el, BP166, 38042
Grenoble, France\\}
\author{W.A. de Heer}
\affiliation{Georgia Institute of Technology, Atlanta, Georgia
30332-0430, USA\\}
%\normalsize{$^\ast$To whom correspondence should be addressed; E-mail:  deheer@physics.gatech.edu.}

%%%%%%%%%%%%%%%%% END OF PREAMBLE %%%%%%%%%%%%%%%%

%%%%%% ABSTRACT %%%%%%
\begin{abstract}
Realization of post-CMOS graphene electronics requires production of semiconducting graphene, which has been a labor-intensive process.\cite{Han2007a,Chen2007b,Han2009,Tapaszto2008,Jiao2009} We present tailoring of silicon carbide crystals via conventional photolithography and microelectronics processing to enable templated graphene growth on $4H$-SiC$\{1\bar{1}0n\}$ $(n\approx8)$ crystal facets rather than the customary \{0001\} %and (000$\bar{1}$) 
planes. This allows self-organized growth of graphene nanoribbons with dimensions defined by those of the facet. Preferential growth is confirmed by Raman spectroscopy and high-resolution transmission electron microscopy (HRTEM) measurements, and electrical characterization of prototypic graphene devices is presented. Fabrication of $>$ 10,000 top-gated graphene transistors on a 0.24 cm$^2$ SiC chip demonstrates scalability of this process and represents the highest density of graphene devices reported to date.
\end{abstract}

\vspace*{4ex}
\pacs{73.21.Ac, 71.20.Tx, 61.48.De, 61.05.cm, 79.60.-i}
\keywords{Graphene, bottom-up, self-organization, SiC, Silicon carbide, nanoribbon, facet, photolithography}
\maketitle
%\newpage

%\textbf{Realization of post-CMOS graphene electronics requires production of semiconducting graphene, which has been a labor-intensive process.\cite{Han2007a,Chen2007b,Han2009,Tapaszto2008,Jiao2009} We present tailoring of silicon carbide crystals via conventional photolithography and microelectronics processing to enable templated graphene growth on 4\textbf{\textit{H}}-SiC\{1$\bar{\textbf{1}}$0\textit{\textbf{n}}\} ($\textit{\textbf{n}}\approx\textbf{8}$) crystal facets rather than the customary \{0001\} %and (000$\bar{1}$) 
%planes. This allows self-organized growth of graphene nanoribbons with dimensions defined by those of the facet. Preferential growth is confirmed by Raman spectroscopy and high-resolution transmission electron microscopy (HRTEM) measurements, and electrical characterization of prototypic graphene devices is presented. Fabrication of $>$ 10,000 top-gated graphene transistors on a 0.24 cm$^2$ SiC chip demonstrates scalability of this process and represents the highest density of graphene devices reported to date.}

%%%%%% INTRODUCTION %%%%%%
Epitaxial graphene on SiC is an exciting new electronic material that presents the possibility of room temperature ballistic devices.\cite{Berger2004,Berger2006,deHeer2007} Extremely high carrier mobilities, exceeding \unit[250,000]{cm$^2$/(V $\cdot$ s)} at room temperature, have been observed,\cite{Orlita2008} and epitaxial graphene has been shown to be compatible with traditional top-down processing techniques.\cite{Kedzierski2008} Its linear semi-metallic band structure\cite{SprinklePRL2009} allows ambipolar tuning of conduction and direct application to RF devices,\cite{Moon2009} while other devices require modification of the graphene sheet to induce a band gap. The latter has received significant attention, and has been accomplished by selective chemical treatment,\cite{Wu2008a,Bekyarova2009} and nanoribbon fabrication,\cite{Han2007a,Chen2007b,Han2009,Tapaszto2008,Jiao2009} which results in a band gap inversely proportional to ribbon width.\cite{Nakada1996,Fujita1996,Wakabayashi2001} The nanoribbon approach is promising in that it has shown large gating effects, but production methods documented thus far, including e-beam lithography\cite{Han2007a,Chen2007b,Han2009} and other approaches,\cite{Tapaszto2008,Jiao2009} are too slow and not controllable enough for technological application. 

\begin{figure}
%\centering
\includegraphics[width=10cm,clip]{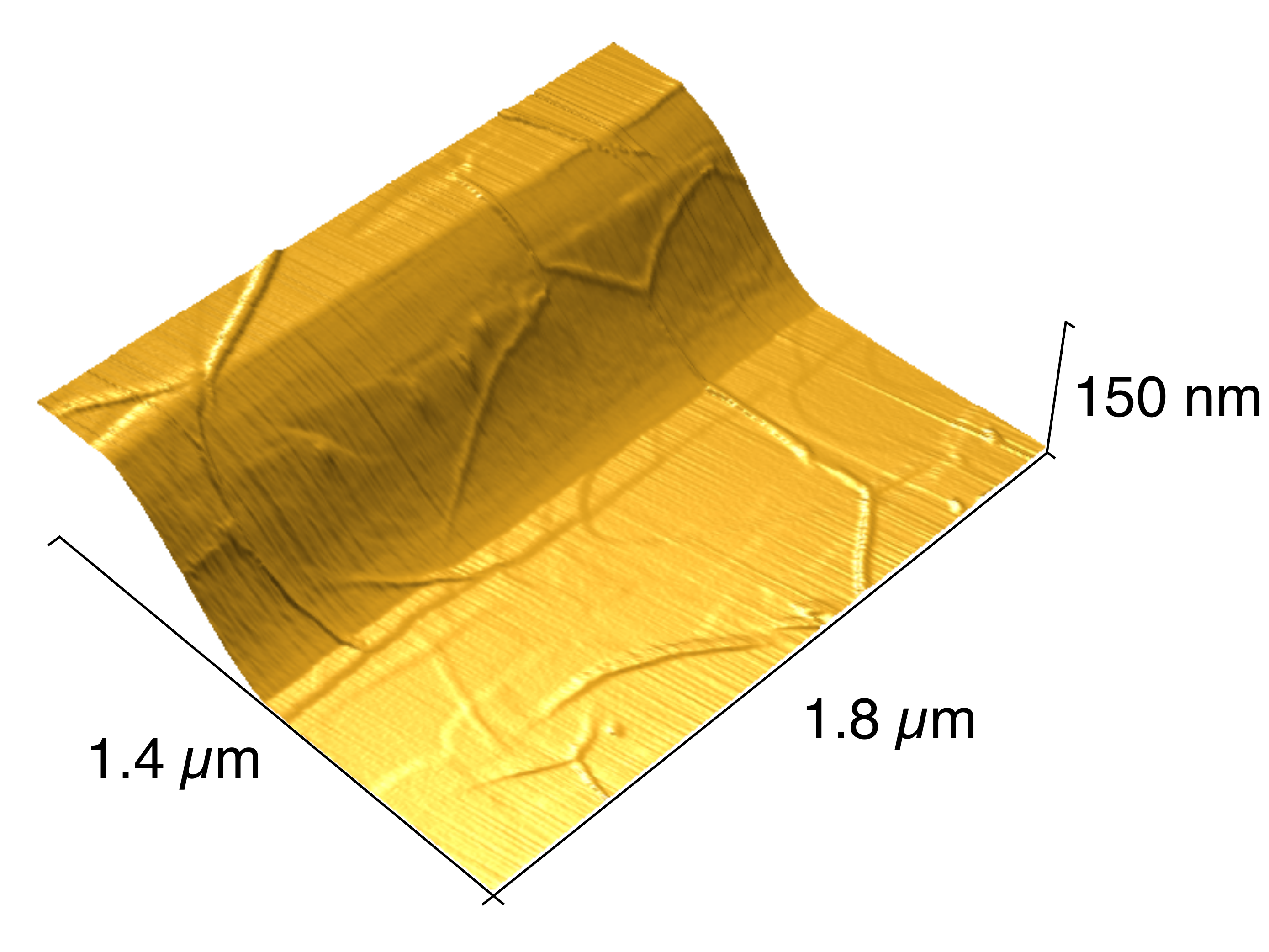}
\caption{AFM scan demonstrating continuity of multilayer epitaxial graphene over a $\sim\!\unit[145]{nm}$ step on the SiC(000$\bar{1}$) plane. Raised pleats are caused by relative contraction of SiC as the sample cools from elevated growth temperatures,\cite{Hass2008b} and their traversal of the step clearly indicates continuity of graphene over the step.} \label{F:AFM}
\end{figure}

Morphology of epitaxial graphene on SiC is highly influenced by the structure of the underlying substrate. In well-controlled growth conditions,\cite{Berger2006,deHeer2007,Hass2008b,Emtsev2009} nominally on-axis SiC retains an ordered terrace structure that originates in the miscut angle of the SiC wafer. Many interpret these steps as being problematic,\cite{Robinson2010} but scanning tunneling microscopy measurements have consistently observed that the graphene lattice is continuous over such steps. This is true of $\unit[0.5]{nm}$ $4H$-SiC half-unit cell steps and few-nm steps where step bunching occurs.\cite{Berger2004,Hass2008c}  We find [see Fig.~\ref{F:AFM}] that this remains true at a much larger scale; graphene pleats typically present on the SiC(000$\bar{1}$) plane\cite{Hass2008b} are observed traversing prepared SiC steps as large as $\unit[150]{nm}$, indicating continuity of graphene over the step. These observations explain transport measurements in which underlying SiC steps appear to have little effect on mobility\cite{Berger2006,Wu2009a} or observation of the quantum Hall effect,\cite{Wu2009a} and suggest exploitation of the effect to produce nanoribbons by novel fabrication methods. 

It has long been known\cite{Kong1988} that SiC\{0001\} surfaces exhibit step bunching in various environments. Recent systematic studies have found a greater propensity for step bunching on the (0001) face,\cite{Syvaejaervi2002,Nakajima2005,Nie2008,Borovikov2009} with vicinal miscuts toward $\langle1\bar{1}00\rangle$ 
displaying bunching of parallel steps into $(1\bar{1}0n)$ ``nanofacets" up to $4-5$ unit cells in height, oriented at an angle of $\sim25^{\circ}$ to the basal plane.\cite{Nakajima2005,Nie2008} It has been suggested that such nanofacet formation may be understood as a minimization of surface free energy.\cite{Nakagawa2003,Nakajima2005} 
Steps perpendicular to the directions $\langle1\bar{1}00\rangle$ are strongly favored on (0001), such that steps formed macroscopically perpendicular to $\langle11\bar{2}0\rangle$ are microscopically zigzagged, with segments perpendicular to $\langle1\bar{1}00\rangle$ directions.\cite{Syvaejaervi2002,Nakajima2005,Nie2008,Borovikov2009}
The $(000\bar{1})$ face, by contrast, seems to form steps at almost any orientation.\cite{Nie2008} Step-bunched nanofaceting has not been observed there previously,\cite{Nie2008,Borovikov2009} but we show in Fig.~\ref{F:AFM} that a $(1\bar{1}0n)$ facet is induced by pre-processing (see below). These results are qualitatively true of both $6H$- and $4H$-SiC polytypes.\cite{Nie2008} 
It is perhaps expected that graphene growth should proceed first on facets, given the lesser bonding of Si atoms, and this has been observed on etching-induced $(1\bar{1}0n)$
\cite{Robinson2010} and $(11\bar{2}n)$\cite{Tanaka2010} nanofacets.

\begin{figure}
%\centering
   	 \subfloat
    	{
	\includegraphics[width=7.5cm,clip]{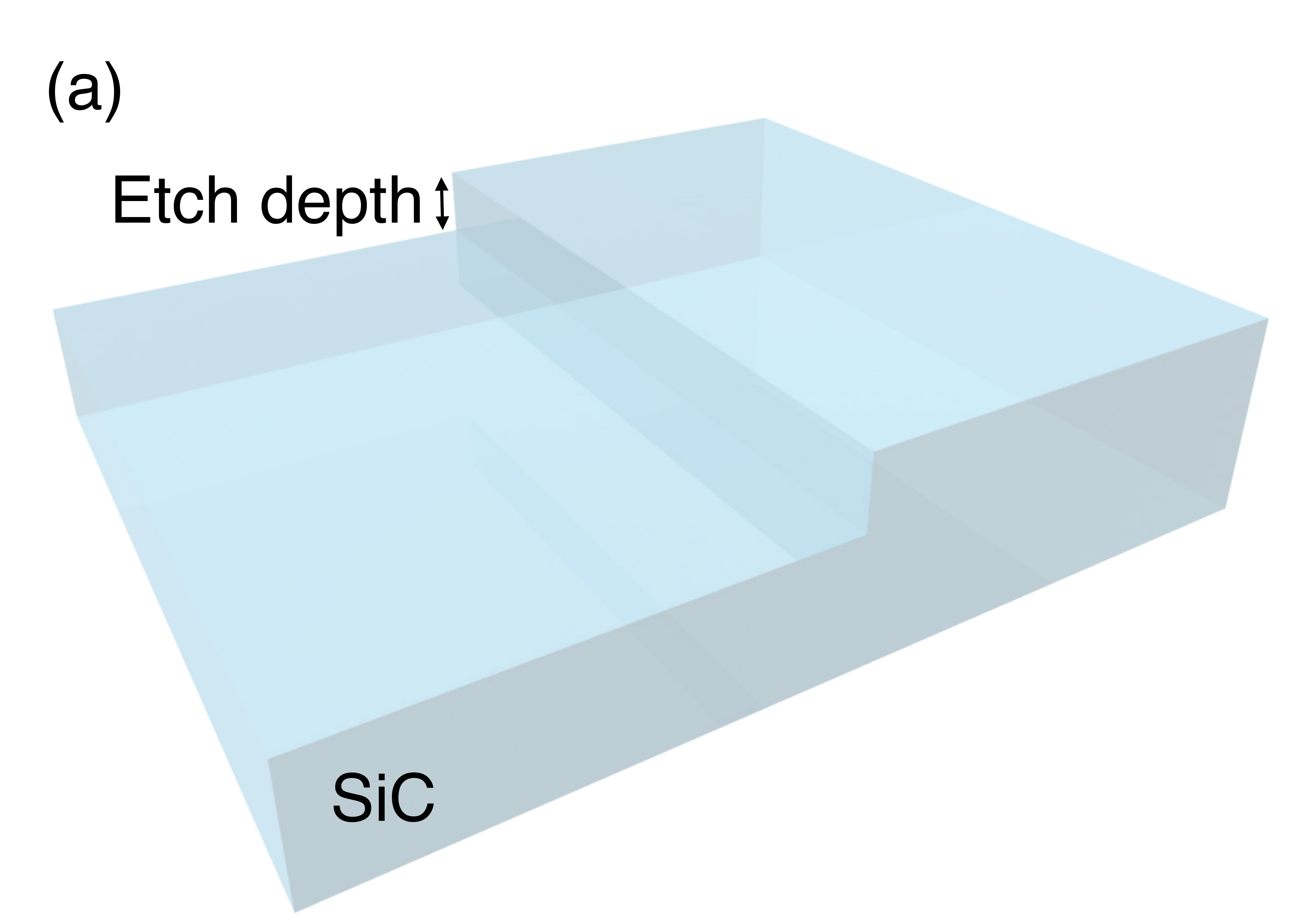}
	}
   	 \subfloat
    	{
	\includegraphics[width=7.5cm,clip]{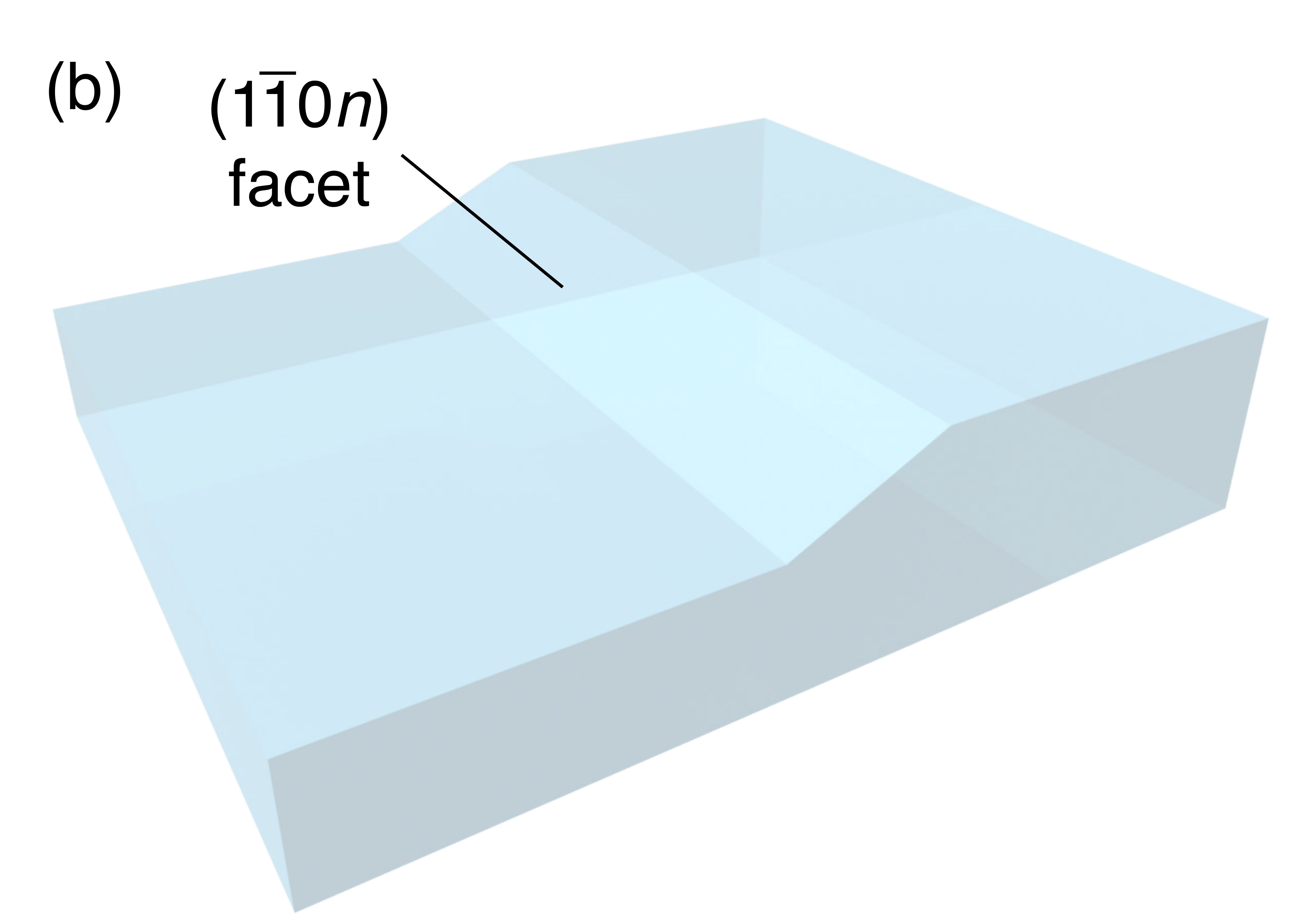}
	}
	\\
   	 \subfloat
    	{
	\includegraphics[width=7.5cm,clip]{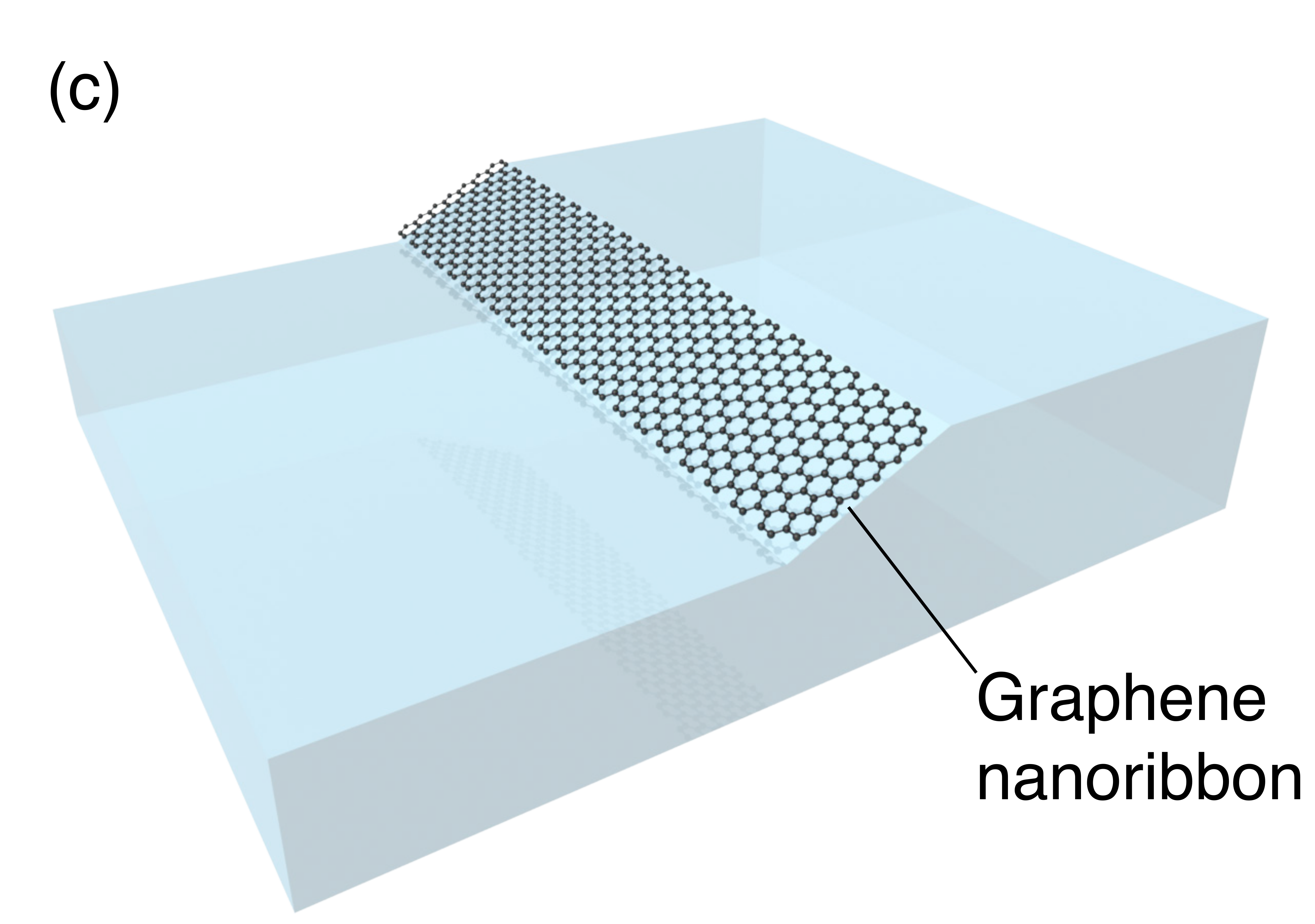}
	}
   	 \subfloat
    	{
	\includegraphics[width=7.5cm,clip]{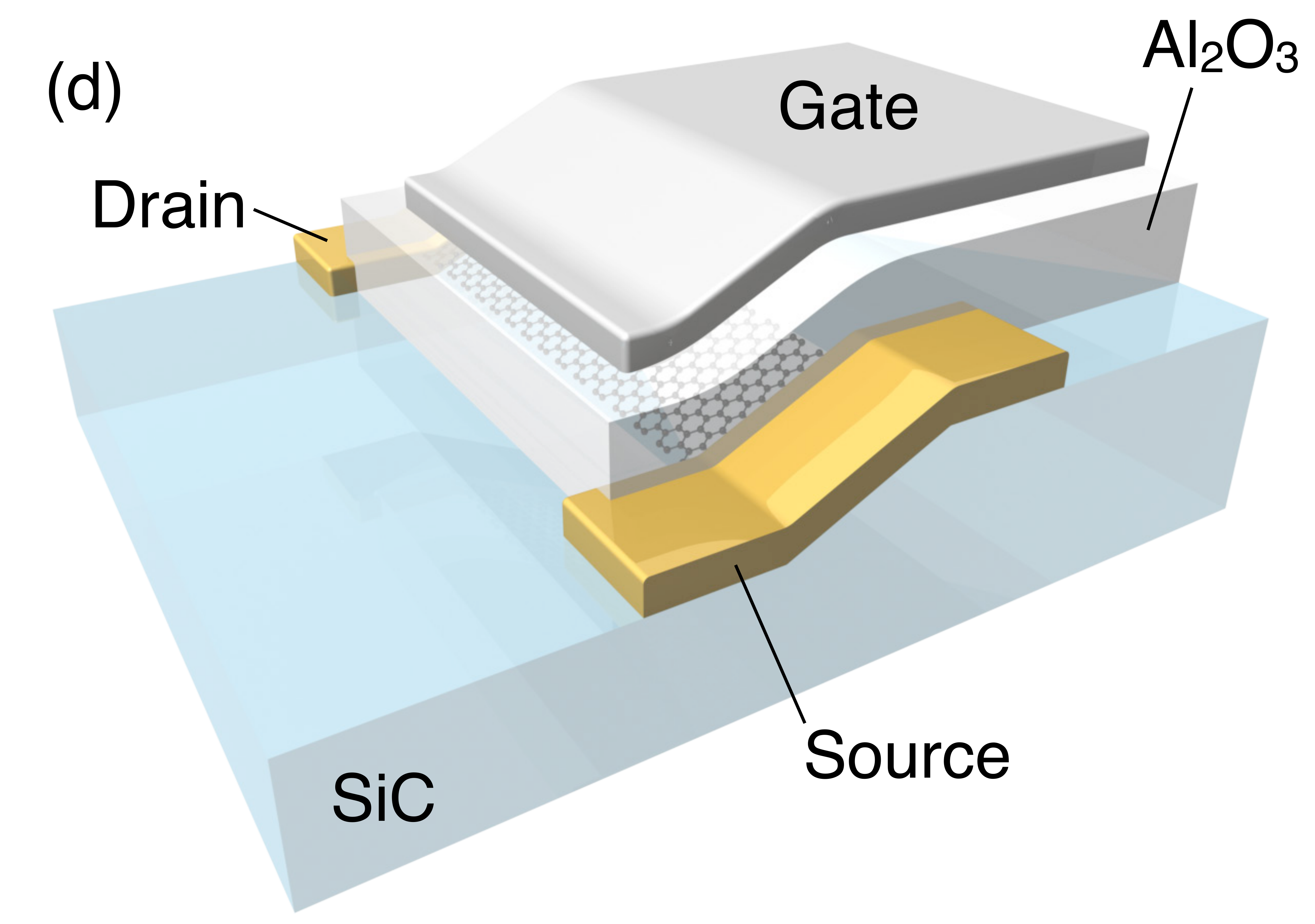}
	}
\caption{Process for tailoring of the SiC crystal for selective graphene growth and device fabrication. (a) nm-scale step is etched into SiC crystal by fluorine-based RIE. (b) Crystal is heated to 1200 -- 1300 $^{\circ}$C (at low vacuum), inducing step flow and relaxation to the $(1\bar{1}0n)$ facet. (c) Upon further heating to $\sim\!1450 ^{\circ}$C, self-organized graphene nanoribbon forms on the facet. (d) Complete device with source and drain contacts, graphene nanoribbon channel, Al$_2$O$_3$ gate dielectric, and metal top gate, as pictured in Fig.~\ref{fig:2pt}.} \label{F:process}
\end{figure}

We propose exertion of control over the natural step bunching mechanism to prepare a crystal facet for self-organized graphene growth [see Fig.~\ref{F:process}]. 
Given the discussion above, the best choice for this purpose may be $(1\bar{1}0n)$.
Controlled facets are achieved by photolithographic definition of Ni lines on a SiC substrate perpendicular to the $\langle1\bar{1}00\rangle$ direction; these lines are transferred into the SiC by a fluorine-based reactive ion etch (RIE), which, while relatively simple technologically, allows nm-precision in the etch depth. As depicted in Fig.~\ref{F:process}, it is the etch depth that ultimately defines the width of nanoribbons prepared. $\unit[15]{nm}$ etch depths were readily achieved in this work, which resulting ribbon width ($\sim\!\unit[30]{nm}$) is sufficiently narrow to result in a sizable band gap at room temperature. Much narrower widths should be reachable with minimal effort. We have nevertheless chosen to focus here on $\sim\!\unit[200]{nm}$ ribbon widths that allow convincing demonstration of the concept via accessibility to characterization probes including Raman spectroscopy, yet narrow enough to exhibit a band gap at low temperature. After removal of the Ni mask and cleaning, the crystal is heated to elevated temperatures (1200 -- 1300 $^{\circ}$C) at intermediate vacuum (10$^{-4}$ Torr) for 30 min., inducing SiC step flow; the abrupt step relaxes to a $(1\bar{1}0n)$ facet, and the temperature is elevated to $> 1450~^{\circ}$C for graphene growth as described previously.\cite{Berger2004,Berger2006,Hass2008b} 

%%%%%% RESULTS %%%%%%

\begin{figure}
%    \centering
    \subfloat
    {
        \includegraphics[width=10cm,clip]{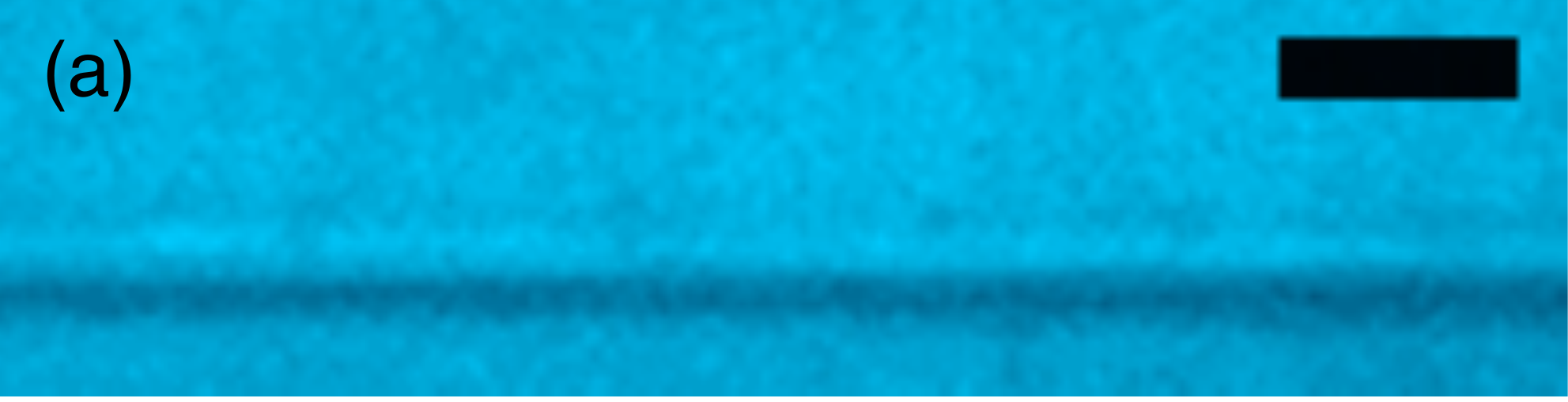}
        \label{fig:raman-sel-opt}
    }
    \\
    \subfloat
    {
        \includegraphics[width=10cm,clip]{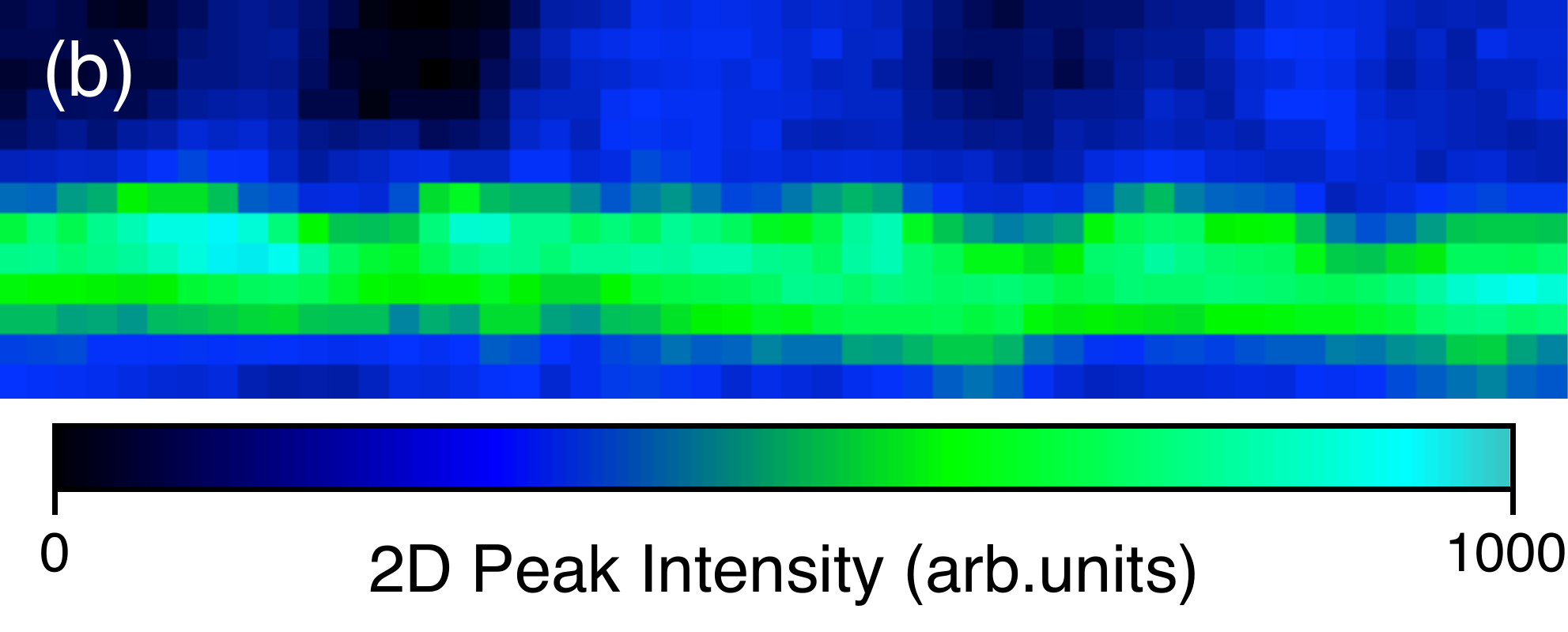}
        \label{fig:raman-sel}
    }
    \\
%    \subfloat
%    {
%        \includegraphics[width=10cm,clip]{colorbar.pdf}
%        \label{fig:raman-sel-scale}
%    }
    \subfloat
    {
        \includegraphics[width=10cm,clip]{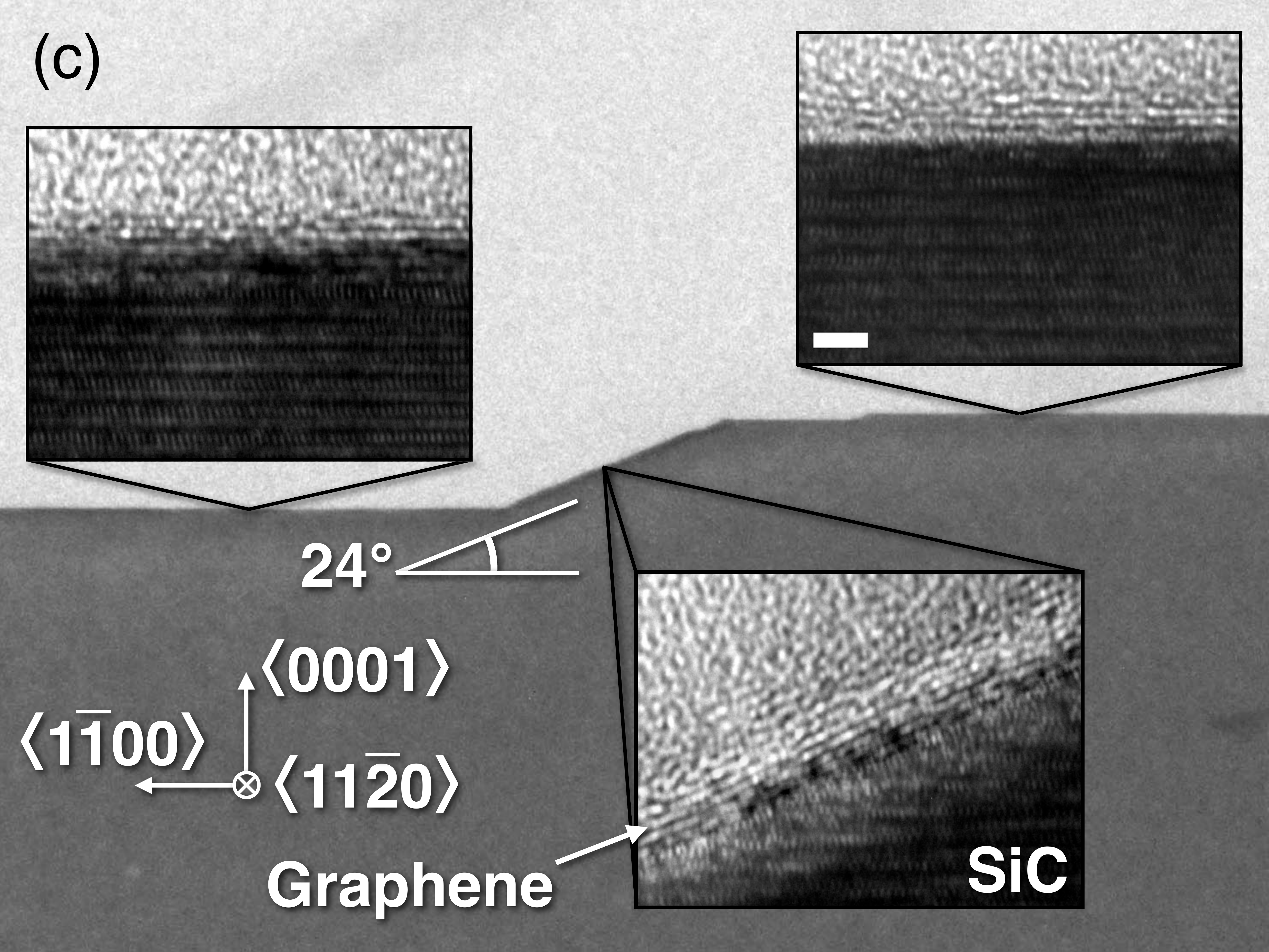}
        \label{fig:TEM}
    }
    \caption{(a) Optical micrograph of a pre-patterned  $\sim\!\unit[100]{nm}$ step on the SiC(0001) face following graphene growth. Scale bar is \unit[2]{$\mu$m}. (b) Raman map ($\sim\!\unit[1]{\mu m}$ lat. res., \unit[0.25]{$\mu$m} grid) of the 2D peak intensity at this location indicates preferential graphene growth on the $(1\bar{1}0n)$ facet. (c) HRTEM cross-sectional images of a similar step confirm preferential growth on the $(1\bar{1}0n)$ facet. Scale bar is 2 nm, and applies to all insets.}
    \label{fig:raman-selective}
\end{figure}

%%% RAMAN/TEM %%%
Preferential growth on the facet is confirmed by Raman mapping as shown in Fig.~\ref{fig:raman-sel}. The intensity of the 2D Raman band (\unit[2700]{cm$^{-1}$}) characteristic of graphene is mapped over a $\sim\!\unit[100]{nm}$ SiC step and adjacent (0001) faces. Little to no intensity is observed on the horizontal surfaces, but significant intensity is seen at the step edge, indicating presence of graphene there. Note that the lateral resolution of the Raman instrument, at $\sim\!\unit[1]{\mu m}$, is much larger than the facet width and the mapping grid spacing. Optimization of growth temperature and time will foreseeably enhance selectivity. 

Cross-sectional HRTEM images [see Fig.~\ref{fig:TEM}] also evidence preferential growth. Multiple layers of graphene are observed on the facet, with only partial layers on the horizontal (0001) plane. The facet angle observed, $24^{\circ}$, is in agreement with AFM measurements (not shown) of $24\!-\!28^{\circ}$ across multiple samples and locations, corresponding to the high-index SiC facet $(1\bar{1}0n)$, where $n\approx8$. The precise facet obtained is dependent on processing temperature.\cite{Syvaejaervi2002}

\begin{figure}
    \subfloat
    {
%    \capstart
        \label{F:Raman-opt}
        \includegraphics[width=7.5cm,clip]{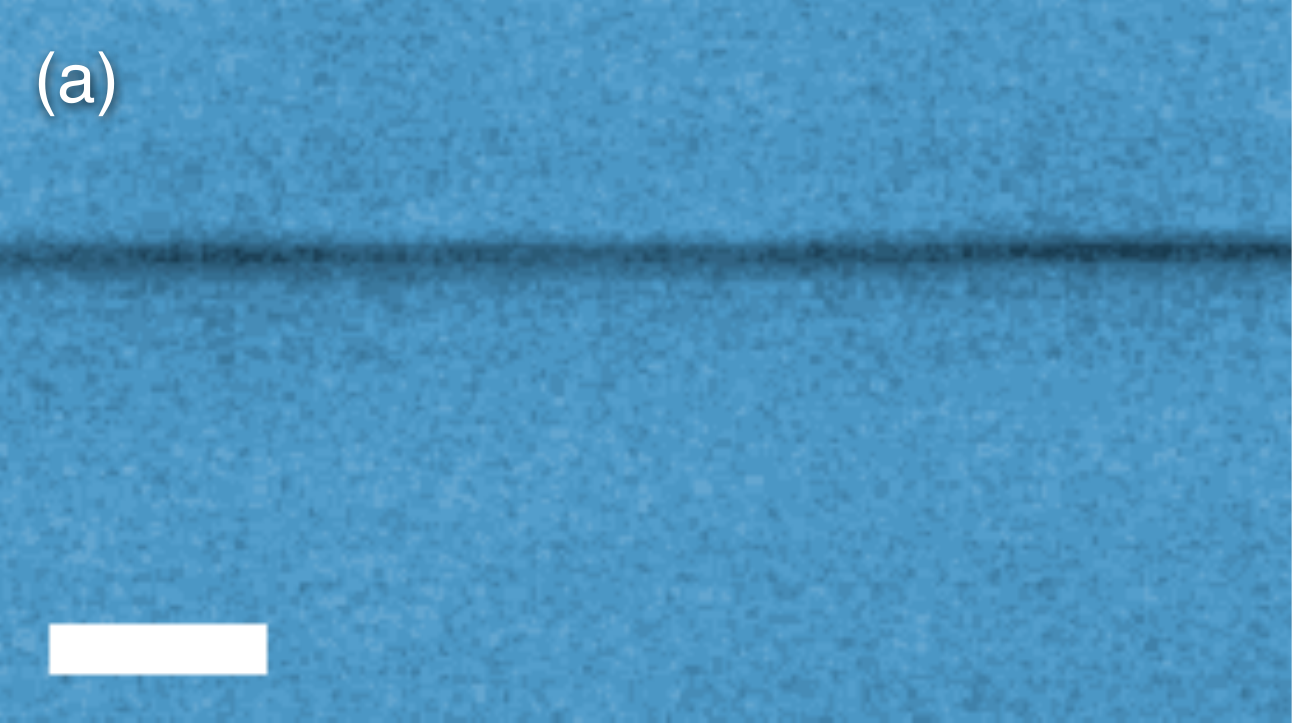}
%        \caption{}
    }
    \\
    \subfloat
    {
%        \capstart
        \label{F:Raman}
        \includegraphics[width=7.5cm,clip]{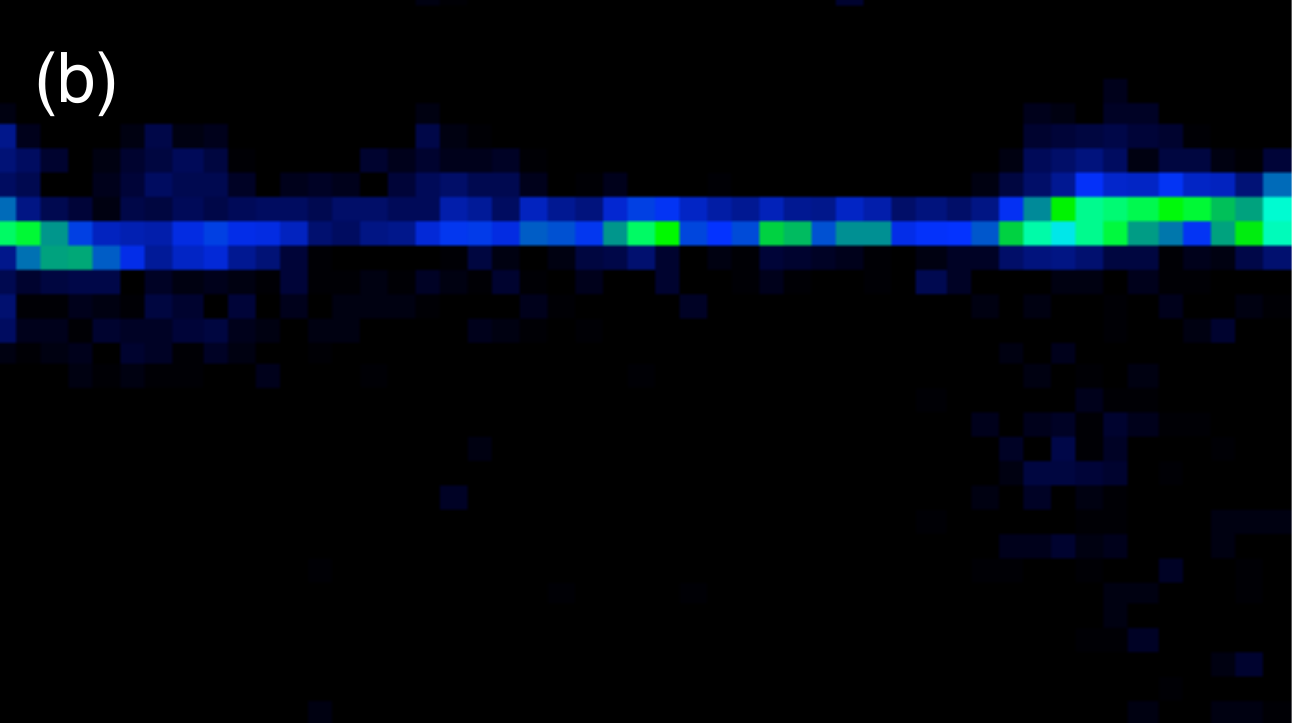}
        
    }
    \\
%    \subfloat
%    {
%        \includegraphics[width=10cm,clip]{4ptinset.jpg}
%        \label{F:4pt-opt}
%    }
%    \subfloat
%    {
%%    \capstart
%        \label{F:4pt-RT}
%        \includegraphics[width=7.5cm,clip]{fig2c.pdf}
%    }
    \subfloat
    {
%    \capstart
        \label{F:4pt-transport}
        \includegraphics[width=7.5cm,clip]{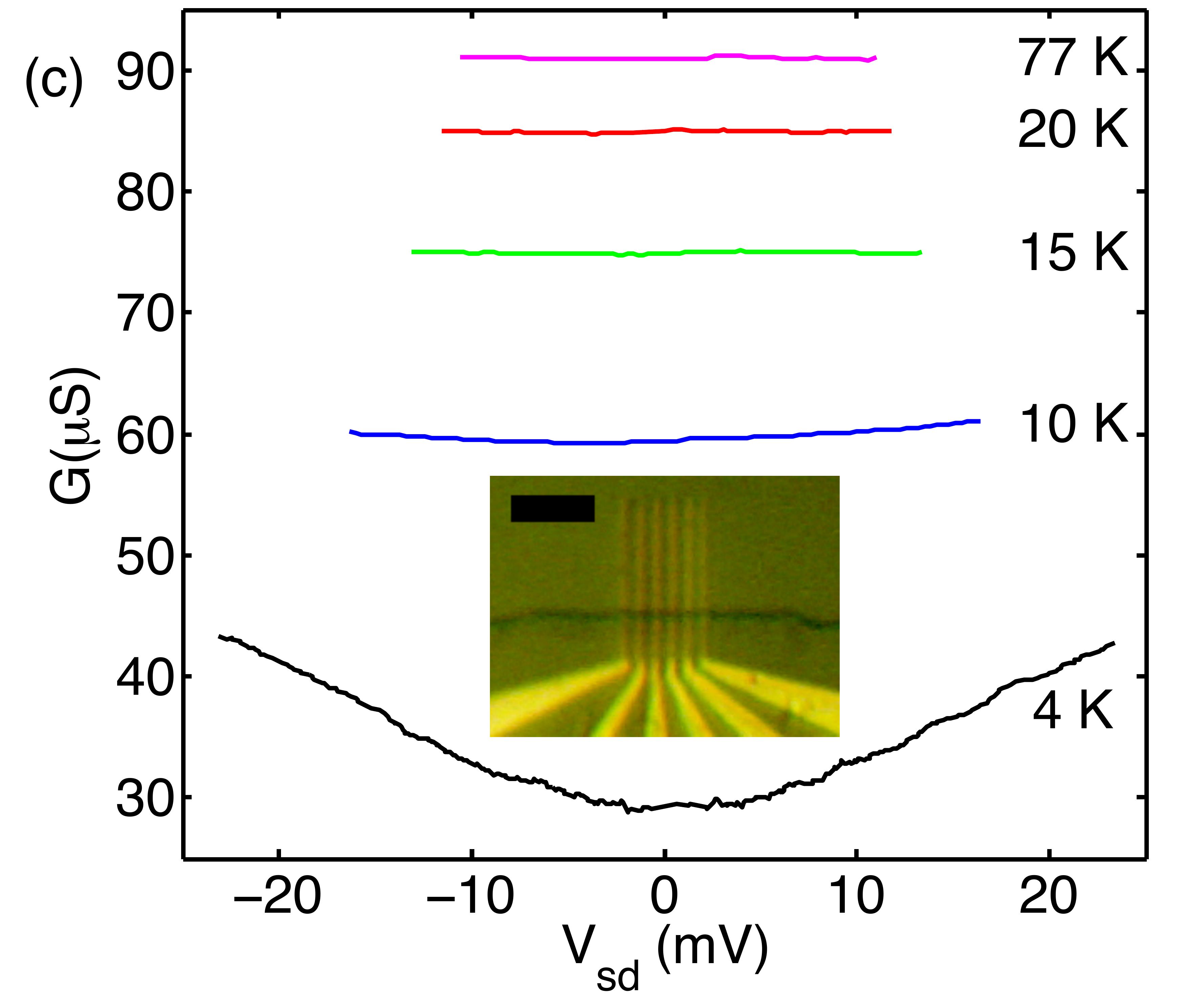}
    }
    \caption{(a) Optical micrograph of  $(1\bar{1}0n)$ facet with graphene ribbon. Scale bar is \unit[2]{$\mu$m}. (b) Raman map ($\sim\!\unit[1]{\mu m}$ lat. res., \unit[0.5]{$\mu$m} grid) of graphene 2D intensity at this location indicates presence of graphene essentially exclusively on the facet. (c) Conductance vs. source-drain voltage as a function of temperature. Band gap is observed at 4K as expected for a ribbon of this width ($\unit[200]{nm}$). Inset: optical micrograph of test structure. Scale bar is \unit[5]{$\mu$m}.}
    \label{fig:4pt}
\end{figure}

Ribbon samples were prepared for electrical measurement by exposure to an extremely short directional O$_2$ RIE to remove any graphene fragments from the horizontal (0001) surface. This was verified by Raman mapping, as shown in Fig.~\ref{F:Raman}. Extensive electrical probing confirmed lack of measurable conductivity on the horizontal surfaces. Note that this processing step will very likely become unnecessary with refinement of preferential growth parameters. Metal contacts were deposited for four-terminal measurements without gate and two-terminal measurements with top gate.  In the latter case, an Al$_2$O$_3$ dielectric was deposited by atomic layer deposition (ALD) followed by deposition of the metal gate [see Fig.~\ref{F:process}].

Measurement of four-terminal test structures [see inset, Fig.~\ref{F:4pt-transport}] yielded sheet resistances of \unit[180-500]{$\Omega$/sq.}, values typically observed in as-grown planar graphene,\cite{deHeer2007} suggesting that the quality of these ribbons does not differ dramatically from that of the planar material. Fig.~\ref{F:4pt-transport} shows a series of conductance vs. source-drain voltage curves taken between 77 K and 4 K. The behavior is metallic at high temperatures, but quantum confinement is clearly manifested in the nonlinearity observed at 4 K, indicating presence of a small band gap, as expected of this $\sim$200 nm ribbon and in agreement with previous reports.\cite{Berger2006} 

\begin{figure}
%    \centering
%    \\
    \subfloat
    {
        \includegraphics[width=10cm,clip]{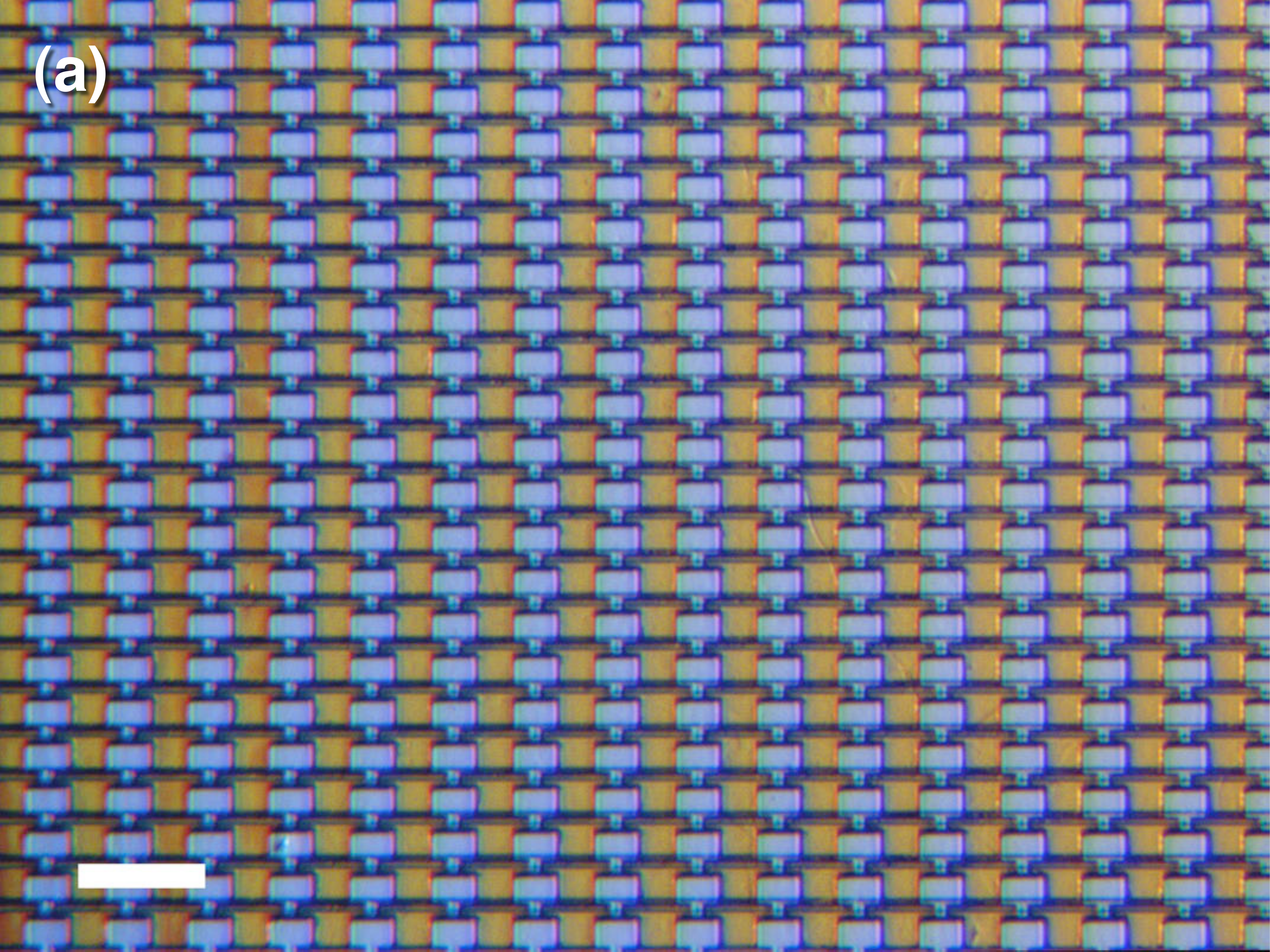}
        \label{fig:2ptoptical}
    }
    \\
    \subfloat
    {
        \includegraphics[width=10cm,clip]{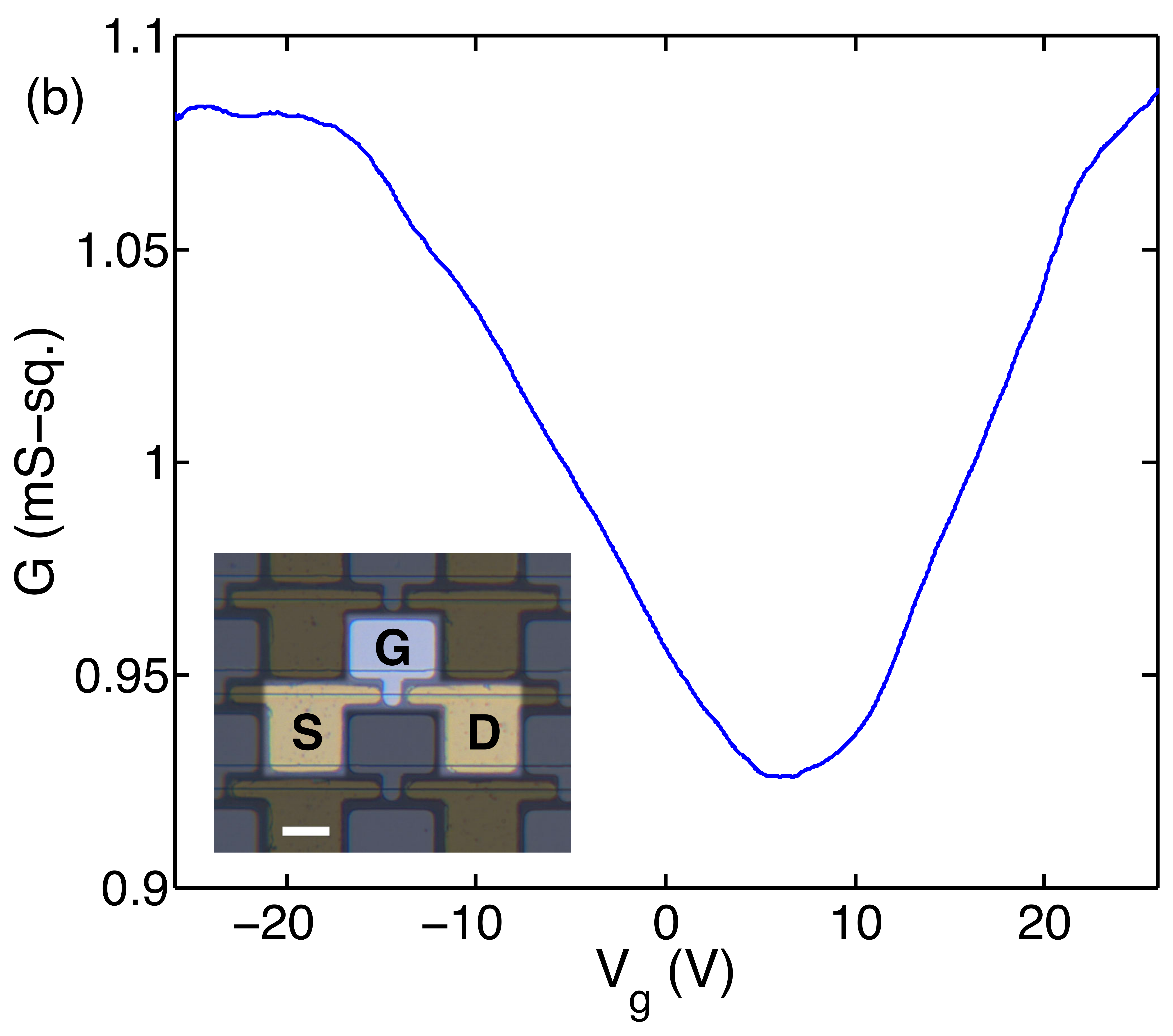}
        \label{fig:2ptIVg}
    }
    \caption{(a) Graphene transistor array with density 40,000 devices per cm$^2$. Scale bar is \unit[100]{$\mu$m}. (b) Room temperature ambipolar gating effect: conductance $G$ vs. gate voltage $V_{\text{sd}}$. Inset: an individual FET consisting of source (S), drain (D), graphene channel, and gate (G). Scale bar is \unit[20]{$\mu$m}.}
    \label{fig:2pt}
\end{figure}

%% and 2-pt.: gate sweep
Photolithographic processing allows fabrication of a large number of devices at higher density. An array of top-gated graphene transistors prepared on the $(000\bar{1})$ face of a \unit[$4\times6$]{mm} SiC chip with SiC etch depth $\sim\!\unit[100]{nm}$ is shown in Fig.~\ref{fig:2ptIVg}. A single device (source, drain, channel, and gate, as illustrated in Fig.~\ref{F:process}) occupies a $35\times\unit[65]{\mu m}$ area, so the $\unit[0.24]{cm^2}$ chip accommodates more than 10,000 transistors. This density was limited primarily by the size of the probe tips used for electrical measurement, but it is, to our knowledge, the highest density of graphene transistors achieved to date. The room temperature gating effect is plotted in Fig.~\ref{fig:2ptIVg}. While transport characteristics of the various devices are reasonably consistent, there is work to be done in improving the selective growth and patterning. Two imminent changes will dramatically enhance this effect: reducing the growth temperature and time to in turn reduce the number of graphene layers on the facet, significantly improving gating efficiency, and reducing the SiC etch depth to narrow the ribbon, inducing a band gap at room temperature.

%%%%%% CONCLUSION %%%%%%
It should be noted that there are likely fundamental differences in the graphene growth among the possible SiC facets, analogous to the dramatic differences in growth speed and layer orientation observed on the (0001) and $(000\bar{1})$ faces,\cite{deHeer2007,Hass2008b} and the $(1\bar{1}0n)$ facet chosen here is possibly not the most desirable in terms of selectivity and quality of graphene produced. This is particularly true of facets prepared on the $(000\bar{1})$ surface, where there is apparently more freedom in facet choice. This is a topic of ongoing research, but the directions are clear.
%edges annealed rather than cut

These results show that graphene growth on non-traditional crystal faces is viable and incredibly useful in device fabrication, particularly for production of nanoribbons on a large scale, and fabrication of   graphene transistors at a density greater than 40,000 per \unit{cm$^{2}$} represents a milestone in the development of graphene electronics. Refinement of this approach appears imminent, as ribbon width is reduced, facet selection and selective graphene growth is optimized, and damage to ribbon edges by violent cutting processes is therefore eliminated. Pre-patterning of the SiC substrate is, in general, a new and promising direction in the development of epitaxial graphene electronics, as more complex structures and applications are readily envisaged. It is furthermore an important coalescence of top-down and bottom-up lithographies.

%%%%%% Materials and METHODS %%%%%%
\begin{center}\textbf{Methods}\end{center}

Substrates were nominally on-axis research-grade semi-insulating $4H$-SiC from Cree, Inc. Arrays of Ni lines were defined on the (0001) or (000$\bar{1}$) SiC crystal face by a standard photolithographic lift-off process, and transferred into the SiC by a 43\% SF$_6$/23\% O$_2$/33\% Ar RIE operating at \unit[30]{mTorr}. RF power was tuned to give a SiC etch rate of $\unit[8]{\AA/s}$, allowing precisely controllable etching. Ultrasonic treatment in nitric acid removed Ni from the SiC surface, and further cleaning and graphene growth proceeded as described previously.\cite{Berger2004,Berger2006,Hass2008b} O$_2$ RIE operating at 100 mTorr was tuned to give a graphene etch rate of $\sim\!\unit[1]{\AA/s}$, and etch time was several to a maximum of ten seconds. Samples were mounted with $(1\bar{1}0n)$ facet parallel to ion flux. Pd/Au contacts were defined by e-beam or photo-lithographic lift-off. Atomic layer deposition (ALD) of Al$_2$O$_3$ was performed as described by Williams et al.\cite{Williams2007} in a commercial Cambridge Nanosystems Savannah ALD system prior to photolithographic lift-off of an Al top-gate. Inclusion of an NO$_2$ functionalization step as specified by Williams was critical for uniform Al$_2$O$_3$ coverage. Raman mapping was performed with excitation wavelength 532 nm, lateral resolution $\sim\!\unit[1]{\mu m}$, and $\unit[0.25-0.5] {\mu m}$ grid spacing. 2D intensity was taken at 2D maxima near $\unit[2725]{cm^{-1}}$. HRTEM measurements were performed by Evans Analytical Group in Raleigh, NC, USA. %5 nm Pd/$\sim$60 nm Au

%%% NOTES %%%
\textbf{Acknowledgements:} We acknowledge helpful conversations with A. Zangwill, V. Borovikov and F. Ming. 
This research was supported by the W.M. Keck Foundation and the NSF under grant No. DMR-0820382. 

%%%%%%%%%%%%%%%%%%%%%%%%%%%%%%%%%%%%%%%%%%%%%%%%%%%%%%%%%%%%%%%%%%%%%%%%%%%%%%%
%%%%%%%%%%%%%%%%%%%%%%%%%%%%%%%%%%%%%%%%%%%%%%%%%%%%%%%%%%%%%%%%%%%%%%%%%%%%%%%

\bibliography{selforganized-arxiv}

%\newpage
\clearpage

\clearpage

%\newpage

\end{document}